\newcommand*{\qt}{\ln q^{\frac{1}{3}}}

\documentclass[nodate,preprintnumbers,amsmath,amssymb,aps,prd]{revtex4}

\begin{document}


%
%

\title{QUANTUM GEOMETRODYNAMICS WITH INTRINSIC TIME DEVELOPMENT}

\author{Chopin Soo}

\address{Department of Physics, National Cheng Kung University,\\
No. 1, University Rd., Tainan 70101,
Taiwan\\
cpsoo@mail.ncku.edu.tw}


\begin{abstract}
Quantum geometrodynamics with intrinsic time development is presented. Paradigm shift from full space-time covariance to spatial diffeomorphism invariance yields a non-vanishing Hamiltonian, a resolution of the `problem of time', and gauge-invariant temporal ordering in an ever expanding universe. Einstein's general relativity is a particular realization of a wider class of theories; and the framework prompts natural extensions and improvements, with the consequent dominance of Cotton-York potential at early times when the universe was small.
\\
\\
Plenary talk, 2nd LeCosPA Symposium ({\it Everything About Gravity}) Taipei, Dec. 17th. 2015, contribution to the Proceedings
\end{abstract}

\keywords{Geometrodynamics; problem of time; quantum gravity.}


\maketitle
\nopagebreak

\section{Introduction}	

   All of us experience the passage of time. But is time an illusion of our perception, an emergent semi-classical entity, or is it present at the fundamental level even in quantum gravity?  General relativity (GR), Einstein's theory of classical space-time,  ties space and time to (pseudo-)Riemannian geometry.
  But in quantum gravity, space-time is a concept of `limited applicability'\cite{Wheeler}. That semi-classical space-time is emergent begs the question what, if anything at all, plays the role of  `time' in quantum gravity? Wheeler went as far as to claim we have to forgo time-ordering, and to declare `there is no spacetime, there is no time, there is no before, there is no after'\cite{Wheeler}.  But without `time -ordering', how is `causality',  which is requisite in any `sensible physical theory', enforced in quantum gravity? Furthermore, a resolution of the `problem of time' in quantum gravity cannot be deemed complete if it fails to account for the intuitive physical reality of time and does not provide satisfactory correlation between time development in quantum dynamics and the passage of time in classical space-times.

  Wheeler also emphasized it is 3-geometry, rather than 4-geometry, which is fundamental in quantum geometrodynamics. The call to abandon 4-covariance is not new. Simplifications in the Hamiltonian analysis of GR, and the fact that the physical degrees of freedom involve only the spatial metric, lead Dirac to conclude that `four-dimensional symmetry is not a fundamental property of the physical world'\cite{Dirac}. A key obstacle to the viability of GR as a perturbative quantum field theory lies in the conflict between unitarity and space-time general covariance: renormalizability can be attained with higher-order curvature terms, but space-time covariance requires time as well as spatial derivatives of the same (higher) order, thus compromising unitarity. Relinquishing 4-covariance to achieve power-counting renormalizability through modifications of GR with higher-order {\it spatial}, rather than space-time, curvature terms was Horava's bold proposal\cite{Horava}.

  Geometrodynamics bequeathed with positive-definite spatial metric is the simplest consistent framework to implement fundamental commutation relations (CR) predicated on the existence of spacelike hypersurfaces.
In the Arnowitt-Deser-Misner (ADM) description of the space-time metric, $ds^2=-N^2dt^2+q_{ij}\left(dx^i+N^idt\right)\left(dx^j+N^jdt\right)$,
these are labeled as constant-$t$ hypersurfaces\cite{ADM}. Quantum states however do not depend on $t$; in his seminal work, DeWitt observed that the wave function of the universe can depend only on the 3-geometry, and time must be determined intrinsically\cite{DeWitt}. The canonical action of GR may be written as
\begin{equation}
S=\int{\tilde\pi}^{ij}\dot{q}_{ij}d^3xdt-\int\left(NH+N^iH_i\right)d^3xdt + {\rm boundary\,\, term},
\end{equation}
wherein the super-Hamiltonian $H=\frac{2\kappa }{\sqrt{q}}\left[G_{ijkl}{\tilde\pi}^{ij}{\tilde\pi}^{kl}+ {\cal V}(q_{ij})\right]$, and
$H_i=-2q_{ij}\nabla_k{\tilde \pi}^{kj}=0$ is the super-momentum constraint which generates spatial diffeomorphisms of the variables.
The DeWitt supermetric, $G_{ijkl}=\frac{1}{2}\left(q_{ik}q_{jl}+q_{il}q_{jk}\right)-l
q_{ij}q_{kl}$, with deformation parameter $l$, has signature $({\rm sgn}[\frac{1}{3}-l], +,+, + ,+ ,+)$. For Einstein's theory,  $l= \frac{1}{2}$  and the potential ${\cal V}=-\frac{q}{(2\kappa)^2}(R - 2\Lambda_{\rm{eff.}}) $. The symplectic potential decomposes as $\int\tilde{\pi}^{ij}\delta
q_{ij}=\int\bar{\pi}^{ij}\delta\bar{q}_{ij}+{\tilde\pi}\delta\ln
q^{\frac{1}{3}}$. So clean separation of the conjugate pair, $(\ln q^{\frac{1}{3}}, {\tilde\pi})$, consisting of (one-third of) the logarithm of the determinant of the spatial metric and the trace of the momentum, from $({\bar q}_{ij}:=q^{-\frac{1}{3}}q_{ij}, {\bar \pi}^{ij}:=q^{\frac{1}{3}}\bigl(\tilde{\pi}^{ij}-\frac{1}{3}q^{ij}\tilde{\pi})\big)$, the unimodular spatial metric and traceless part of the momentum, allows a deparametrization of the theory wherein $\ln q^{\frac{1}{3}}$ (associated with the negative mode in the DeWitt supermetric) plays the role of the intrinsic time variable when $\beta^2:=l-\frac{1}{3} >0$.

A framework for quantum geometrodynamics without the paradigm of space-time covariance, and equipped with spatial diffeomorphism-invariant physical Hamiltonian, has been advocated in a series of works\cite{Soo_Hoi-Lai,  Niall, ITQG, ITG, NCR}. It resolves `the problem of time' and bridges  the deep divide between quantum mechanics and conventional canonical formulations of quantum gravity with a Schr\"{o}dinger equation which describes first-order evolution and time-ordering in global intrinsic time.

\section{Classical GR and intrinsic time development}

The super-Hamiltonian constraint factorizes in an interesting
way as
\begin{equation}
\frac{\sqrt{q}}{2\kappa}H=G_{ijkl}\tilde{\pi}^{ij}\tilde{\pi}^{kl}+{\cal V}(q_{ij})=-\left(\beta{\tilde\pi}-\bar{H}\right)\left(\beta{\tilde\pi}+\bar{H}\right)=0,
\end{equation}
wherein
$\bar{H}=\sqrt{\bar{G}_{ijkl}\bar{\pi}^{ij}\bar{\pi}^{kl}+{\cal V}(q_{ij})}=\sqrt{\frac{1}{2}\left[\bar{q}_{ik}\bar{q}_{jl}+\bar{q}_{il}\bar{q}_{jk}\right]\bar{\pi}^{ij}\bar{\pi}^{kl}+{\cal V}(q_{ij})}$.
This constrains $ \bar H=\pm\beta{\tilde\pi}$. In Einstein's theory when ${\cal V} =- \frac{q}{(2\kappa)^2}[R - 2\Lambda_{\rm{eff.}} ]$ and $ l=\frac{1}{2}$,  the constraints form a first class algebra; and the lapse function, $N$, which is a Lagrange multiplier, is {\it a priori} an arbitrary function. But the physical meaning of $N$ is revealed {\it a posteriori} by the equations of motion (EOM) and constraints.  In particular, the lapse function is related to $\partial_t\ln q$ and ${\bar H}$ (which assumes the role of Hamiltonian density in intrinsic time geometrodynamics) through
\begin{eqnarray}
\frac{\partial{\qt}(x)}{\partial t} &=\{ \qt(x), \int N(y)H(y)d^3y \}_{\rm P.B.}=\int N(y)\{ \qt(x), H(y)\}_{\rm P.B.} d^3y\nonumber \\
&= -\frac{4}{\sqrt q}{N}(x)\kappa\beta^2\tilde\pi(x)\nonumber \\
&= \mp 4\kappa\beta\underset{\sim}{N}(x){\bar H}(x),
\end{eqnarray}
wherein the Poisson bracket $\{ \ln q^{\frac{1}{3}}(x), {\tilde\pi}(y)\}_{\rm P.B.} =\delta(x-y)$ has been used to arrive at the intermediate step, and the Hamiltonian constraint in the last step (even though $N$ was {\it a priori} arbitrary).
This is the precise form of  the {\it emergent} ADM lapse function $N$ in Einstein's theory.

{\it Dispensing with the Hamiltonian constraint}, the EOM of the physical degrees of freedom (d.o.f.) (which lie in the transverse traceless excitations of $(\bar q_{ij}, {\bar\pi}^{ij}))$ can equivalently be captured by the non-trivial Hamiltonian, $H_{\rm ADM}  := \frac{1}{\beta}\int \frac{\partial\ln q^{\frac{1}{3}}(x)}{\partial t}{\bar H}(x) d^3x$, which generates evolution w.r.t. ADM coordinate time $t$. To wit,
\begin{eqnarray}
{\dot{\bar q}}_{ij}(x) &=& \{ {\bar q}_{ij}(x),  \frac{1}{\beta}\int \frac{\partial\qt(y)}{\partial t}{\bar H}(y)d^3y\}_{\rm P.B.} \nonumber\\
&=&\int \frac{1}{\beta}\frac{\partial\qt(y)}{\partial t}\left\{ {\bar q}_{ij}(x), {\bar H}(y)\right\}_{\rm P.B.} d^3y\nonumber\\
&=& \int \mp\underset{\sim}{N}(y)4\kappa{\bar H}(y)\left\{ {\bar q}_{ij}(x), {\bar H}(y)\right\}_{\rm P.B.} d^3y\nonumber \\
&=&\int \mp\underset{\sim}{N}(y)2\kappa\left\{ {\bar q}_{ij}(x), {\bar H}^2(y)\right\}_{\rm P.B.} d^3y\nonumber\\
&=&\int \mp{N}(y)\left\{ {\bar q}_{ij}(x), H(y)\right\}_{\rm P.B.} d^3y.
\end{eqnarray}
This final step can be reached bearing in mind $\frac{\sqrt{q}}{2\kappa} H =\bar H^2 -\beta^2{\tilde\pi}^2$ and $\tilde\pi$ commutes with $(\bar q_{ij}, {\bar\pi}^{ij})$. This is thus equivalent to the evolution generated through conventional $\int {\mp N}(y)\left\{ {\bar q}_{ij}(x), H(y)\right\}_{\rm P.B.} d^3y$ with the {\it a posteriori} relation (3) between $N, \partial_t\ln q$ and ${\bar H}$.
The $\mp$ sign accompanying $N$ does not affect the resultant classical ADM four-metric which depends on the square of $N$. The EOM for $\bar\pi^{ij}$ can be similarly demonstrated.
Adding $\int N^i(y)H_i(y) d^3y$ to the Hamiltonian merely leads to modification of the EOM by Lie derivatives
of the variables w.r.t. $N^i$, with the resultant lapse function (3) generalized to\footnote{The determinant of the ADM metric is equal to $Nq$, so the lapse is non-vanishing unless there is degeneracy. In, for instance, Minkowski space-time, both the numerator and denominator vanish, but we should consider instead
 de Sitter space-time with compact slicings, and $N$ is well-defined in the limit of vanishing positive cosmological constant.}
\begin{equation}\label{lapse}N =\frac{\sqrt{q}(\partial_t \ln q^{1/3} -\frac{2}{3}\nabla_iN^i)}{4\beta\kappa {\bar H}}.
\end{equation}
In fact this ensures the Hamiltonian constraint is satisfied classically in the form $\frac{K^2}{9} = \frac{4\kappa^2\beta^2}{q}{\bar H}^2$. Thus the  Hamiltonian $H_{ADM}$, {\it which is no longer constrained to vanish}, equivalently captures the physical content and EOM of Einstein's theory. This framework however allows the potential ${\cal V}$ {\it to depart from that of Einstein's theory without leading to inconsistencies in the constraint algebra}.

\section{Global intrinsic time development and temporal order}

While $q$ is a tensor density, the multi-fingered intrinsic time interval,  $\delta\qt =\frac{1}{3}\frac{\delta q}{q}= \frac{q^{ij}}{3}\delta q_{ij}$, is a scalar entity\footnote{In DeWitt's seminal work\cite{DeWitt}, the intrinsic time variable was $q^{\frac{1}{4}}$ instead of $\qt$.}.
Hodge decomposition for any compact Riemannian manifold without boundary yields
\begin{equation}
\delta \ln q^{\frac{1}{3}} =\delta{T}+\nabla_i\delta{Y}^i,
\end{equation} wherein the gauge-invariant part of $\delta\qt$ is $\delta T =\frac{2}{3}\delta \ln V$, which is proportional to the 3-dimensional diffeomorphism invariant (3dDI) logarithmic change in the spatial volume $V$.
This can be seen from $\int (\delta\qt) \sqrt{q} d^3x = \int (\delta{T}+\nabla_i\delta{Y}^i) \sqrt{q} d^3x =(\delta{T})\int \sqrt{q} d^3x =(\delta T)V$ (which implies that $\delta T$ is the average value of $\delta\qt(x)$ over $V$),
 and also $\int (\delta\qt )\sqrt{q} d^3x =\frac{1}{3}\int \frac{(\delta{\sqrt{q}}^2)}{{\sqrt q}^2} \sqrt{q}d^3x =\frac{2}{3}\delta(\int \sqrt{q}d^3x)=\frac{2}{3}\delta V$, thus yielding $(\delta T)V =\frac{2}{3}\delta V$. Being the logarithmic change of a tensor density, $\delta\qt$ also has the distinct advantage (over, for instance, the use of simple scalar fields as intrinsic time) that  the $\nabla_i\delta{Y}^i$ piece in its Hodge decomposition is in fact a Lie derivative of $\qt$ (in particular, ${\mathcal L}_{\delta{{\overrightarrow N}}}\qt = \frac{2}{3}\nabla_i{\delta N^i}$) and can thus be gauged away completely with spatial diffeomorphisms.

 Instead of describing the evolution of quantum states w.r.t.  gauge-dependent multi-fingered time $\qt(x)$, it is eminently more meaningful to ask how the wave function of the universe, $\Psi$, changes w.r.t to 3dDI {\it global} intrinsic time variable $T$.
With $T$ displacing the physically less concrete ADM coordinate time $t$ in $H_{ADM}$, this dynamics is determined  by the Schr\"{o}dinger equation,
 \begin{equation}i\hbar \frac{\delta \Psi}{\delta T} = {H}_{\rm Phys}\Psi, \qquad H_{\rm Phys} := \int \frac{\bar H(x)}{\beta} d^3x ;\end{equation}
wherein $H_{\rm Phys}$ is the physical Hamiltonian generating evolution in global intrinsic time.  This {\it replaces, and circumvents, the naive imposition of the Hamiltonian constraint as a Schwinger-Tomonaga equation with multi-fingered time} $\ln q^{\frac{1}{3}}(x)$, which would have looked like ${i\hbar}\frac{\delta\Psi}{\delta \ln q^{\frac{1}{3}}(x)}\Psi =\frac{\bar H(x)}{\beta}\Psi $  in the metric representation.

The formalism is applicable to the full theory of quantum gravity\cite{ITQG, NCR}, and assumptions of mini-superpsace models have not been invoked. $\beta >0$ is required for the Hamiltonian to be bounded from below, and global intrinsic time increases monotonically with our ever expanding universe.

In this framework dictated by first-order Schr\"{o}dinger evolution
in global intrinsic time, the Hamiltonian ${H}_{\rm Phys}$ is 3dDI provided ${\bar H} = {\sqrt{ {\bar \pi^{ij}}  {\bar G_{ijkl}} {\bar \pi^{kl}} + {\cal V}}}$
is a scalar density of weight one; and Einstein's GR (with its particular value of $\beta$ and ${\cal V}$ ) is a particular realization of this wider class of theories.
 The crucial time development operator can be derived by integrating the Schr\"{o}dinger equation, yielding $\Psi(T) = U(T,T_0)\Psi(T_0)$,
 with $T$-ordered operator $U(T,T_0):={\cal T}\exp\left[-\frac{i}{\hbar}\int^T_{T_0} {H}_{\rm Phys}(T')dT'\right]$, or in the form of a time-ordered Dyson series,
 \begin{eqnarray}
 &U(T,T_o)=I - \frac{i}{\hbar}{\int}^{T}_{T_o}dT_1{H}_{\rm Phys}(T_1)+\nonumber\\&(\frac{-i}{\hbar})^2 \int^{T}_{T_o}dT_2\int^{T_2}_{T_o}dT_1
 {H}_{\rm Phys}(T_2){H}_{Phys}(T_{1})+...+\nonumber\\&(\frac{-i}{\hbar})^n\int^{T}_{T_o}dT_n\int^{T_n}_{T_o}dT_{n-1} ...\int^{T_{2}}_{T_o}dT_1 {H}_{\rm Phys}(T_n){H}_{\rm Phys}(T_{n-1})...{H}_{\rm Phys}(T_1) +...\,.
 \end{eqnarray}
Unitary and diffeomorphism-invariant $U(T,T_0)$  follows if the 3dDI ${H}_{\rm Phys}$ is self-adjoint. Since $dT$  is also unchanged under spatial diffeomorphisms, the temporal ordering  in $U(T, T_0)$ will be reassuringly gauge invariant.
Time-ordered integrability of the  Schr\"{o}dinger equation is feasible without ambiguity because $\delta T$ is `1-dimensional', more precisely, spatially-independent {\it rather than many-fingered}, and $H_{\rm Phys}$ is a spatial integral, rather than a local Hamiltonian density. Moreover, the necessity of `time'-ordering, which underpins the notion of causality, emerges, because quantum fields and the Hamiltonian do not commute at different `times'\footnote{The Hamiltonian in fact depends explicitly upon intrinsic time.}.
What is paramount to causality is not the dimension of time, but the sequence, or time-ordered development, of the physical state.

Wheeler's provocative call to forgo all time-ordering may ring true intuitively, but it does not hold up to deeper scrutiny. While space-time events and their ordering have no primary role and space-time itself is an entity `of limited applicability'  in quantum gravity, the fundamental gauge-invariant entity that can be, and is in fact, time-ordered is the physical quantum state.

\section{Modification of the potential, and asymptotic behavior of the Hamiltonian at early and late intrinsic times}

Paradigm shift from 4-covariance to 3dDI  (or spatial covariance), not only allows a resolution of the `problem of time', but it also permits the introduction of higher spatial derivative terms\cite{Horava} in the potential ${\cal V}$ in ${\bar H}$ to improve convergence without sacrificing unitarity. Since ${\bar H}$ is a square root, this suggests a positive semi-definite weight two potential\cite{Soo_Hoi-Lai, ITQG},
\begin{equation}
\begin{split}
&{\cal V}=[\frac{1}{2}( {\bar q_{ik}}{\bar q_{jl}} + {\bar q_{il}}{\bar q_{jk}} ) +\gamma {\tilde q_{ij}} {\tilde q_{kl}}] {\tilde {\cal W}^{ij} }{\tilde{\cal  W}^{kl}},  \quad  \gamma \geq -\frac{1}{3};\\
&{{\tilde{\cal W}}^{i}_{j} }=[\sqrt q (\Lambda' + a' R)\delta^{i}_{j} + b\hbar\sqrt q {\bar R}^{i}_{j} +g\hbar\tilde C^{i}_{j} ];
\end{split}
\end{equation}\noindent
wherein $R$  and ${\bar R}^i_j$ are respectively the scalar and traceless parts of the spatial Ricci curvature, while ${\tilde C}^i_j$ is the Cotton-York tensor (density) which is third order in spatial derivatives and associated with dimensionless coupling constant $g$. This is compatible with the introduction of higher spatial curvature terms up to ${\tilde C}^i_j{\tilde C}^j_i$ in $H$ for power-counting renormalizability.\cite{Horava}

Hodge decomposition for $\delta \qt$ and its Heisenberg equation of motion lead to $\frac{d}{dT} \qt(x, T)= \frac{\partial}{\partial T}\qt  + \frac{1}{i\hbar}[\qt, H_{\rm Phys}] = 1$; with solution
$\ln[\frac{q(x, T) }{q(x,T_{\rm now})}]=3{(T-T_{\rm now})}$, and $-\infty < T < \infty$. Moreover, $\delta T = \frac{2}{3}\delta \ln V$ i.e. $T-T_{\rm now} = \frac{2}{3}\ln(V/V_{\rm now})$. Explicitly separating out T-dependence from entities (labeled with overline) which depend  only on ${\bar q}_{ij}$ yields
\begin{eqnarray}
\label{INTRINSIC2}
&{\tilde{\cal W}^{i}_{j} }=[\sqrt q (\Lambda' + a' R)\delta^{i}_{j} + b' \sqrt q {\bar R}^{i}_{j} +g\hbar\tilde C^{i}_{j} ]\nonumber\\
&=[\sqrt q (\Lambda' + a' q^{-\frac{1}{3}}{\bar q}^{kl}{\overline R}_{kl})\delta^{i}_{j} + b' {\sqrt q}q^{-\frac{1}{3}}\bar q^{ik}\overline{\bar R}_{kj} +g\hbar\tilde C^{i}_{j} ] + {\sqrt q}(\partial_{i}\ln q \,\,{\rm terms}),
\end{eqnarray}
with $q$-independent Cotton-York tensor density $\tilde C^{i}_{j}$ which is conformally invariant.
The Hamiltonian is explicitly (intrinsic)time-dependent, and not (intrinsic)time-reversal invariant; furthermore, the exponential scaling behavior of $q$ with intrinsic time implies  in the limit $ T -T_{\rm now} \rightarrow -\infty,  V/V_{\rm now} \rightarrow 0 $ (i.e. early times when the universe was very small in volume), the potential $\mathcal{V}$  was dominated by the Cotton-York term; whereas the limit $ T -T_{\rm now} \rightarrow \infty,  V/V_{\rm now} \rightarrow \infty $ (i.e. late times when the universe becomes large) will be dominated by the cosmological constant term, which is compatible with current observations and understanding of our ever expanding universe\footnote{An exemplar of the determination of global intrinsic time interval is its measurement through dimensionless redshift in homogeneous Friedmann-Lemaitre-Robertson-Walker (FLRW) cosmology.}. In the  middle period, curvature and cosmological terms will be comparable in importance.

\section{Momentric variable, and the free theory}

The Poisson brackets for the $(\bar q_{ij}, {\bar\pi}^{ij})$ variables are,
\begin{eqnarray}\label{qT}
&\{{\bar{q}}_{ij}({\bf x}),\bar{q}_{kl}({\bf y})\}_{\rm P.B.} =0, \quad
\{\bar{q}_{kl}({\bf x}),{\bar{\pi}}^{ij}({\bf y})\}_{\rm P.B.}= P^{ij}_{kl}\,\delta({\bf x}-{\bf y}),\nonumber\\
&\{{\bar{\pi}}^{ij}({\bf x}),{\bar{\pi}}^{kl}({\bf y})\}_{\rm P.B.}=\frac{1}{3}({\bar q}^{kl}{\bar{\pi}}^{ij} -{\bar q}^{ij}{\bar{\pi}}^{kl})\delta({\bf x}-{\bf y});
\end{eqnarray}
with $P^{ij}_{kl} :=  \frac{1}{2}(\delta^i_k\delta^j_l + \delta^i_l\delta^j_k) - \frac{1}{3}\bar{q}^{ij}\bar{q}_{kl}$ denoting the traceless projection operator. This set is not thus strictly canonical, and difficulties in implementing ${\bar\pi}^{ij}$ as self-adjoint traceless operator in the metric representation lead us to summon the momentric variable (first introduced by Klauder\cite{Klauder}) which is classically $\bar \pi^{i}_{j} = \bar q_{jm}\bar \pi^{im}$. The fundamental CR of the traceless momentric and unimodular part of the spatial metric are
\begin{equation}
\label{RELATIONS}
\begin{split}
&[ \bar q_{ij}(x), \bar q_{kl}(y)]=0,
\quad [\bar q_{ij}(x), {\bar{\pi}}^{k}_{l}(y)]= i\hbar\bar{E}^k_{l(ij)}\delta(x-y),\\
&[ {\bar{\pi}}^{i}_{j}(x), {\bar{\pi}}^{k}_{l}(y)]= \frac{i\hbar}{2}\bigl(\delta^k_j{\bar{\pi}}^i_l-\delta^i_l{\bar{\pi}}^k_j\bigr)\delta(x-y);
\end{split}
\end{equation}\noindent
wherein $\bar{E}^i_{j(mn)}=\frac{1}{2}\bigl(\delta^i_m\overline{q}_{jn}+\delta^i_n\overline{q}_{jm}\bigr)-\frac{1}{3}\delta^i_j\overline{q}_{mn}$
is the vielbein for the supermetric ${\bar G}_{ijkl} =  \bar{E}^m_{n(ij)}\bar{E}^n_{m(kl)}$.
Quantum mechanically, the momentric operators and CR can be explicitly realized  in the metric representation by
\begin{equation}
{\bar{ \pi}}^{i}_{j}(x)=\frac{\hbar}{i}\bar{E}^i_{j(mn)}(x)\frac{\delta}{\delta \bar q_{mn}(x)} =\frac{\hbar}{i}\frac{\delta}{\delta \bar q_{mn}(x)}\bar{E}^i_{j(mn)}(x)={\bar{ \pi}}^{\dagger i}_{j}(x)
\end{equation}\noindent
which are self-adjoint on account of $[\frac{\delta}{\delta\bar{q}_{mn}(x)},\bar{E}^i_{j(mn)}(x)]=0$.
These eight momentric variables generate $SL(3,R)$ transformations which preserve positivity and unimodularity of $\bar{q}_{ij}$. Explicitly, $U^\dagger(\alpha) {\bar q}_{kl}(x) U(\alpha) = (e^{\frac{\alpha(x)}{2}})^m_k {\bar q}_{mn}(x) (e^{\frac{\alpha(x)}{2}})^n_l$,  wherein $ U(\alpha)=e^{-\frac{i}{\hbar}\int \alpha^i_j {\bar \pi}^j_i d^3y}$. Furthermore,  they generate, by themselves, at each spatial point, an $SU(3)$ algebra. In fact, defining $T^{A}(x):= \frac{1}{\hbar\delta(0)}(\lambda^{A})^{j}_{i}{\bar \pi}^{i}_{j}(x)$ with Gell-Mann matrices $\lambda^{A=1,...,8}$ lead to
 \begin{equation}[T^{A}(x),T^{B}(y)]= i{f}^{AB}\,_CT^{C}\frac{\delta (x-y)}{\delta(0)},\end{equation}
 with $SU(3)$ structure constants $f^{AB}\,_C$\cite{Gell-Mann}. It is noteworthy that in the absence of the potential ${\cal V}$ the
 the free theory is characterized by $SU(3)$ invariance generated by the momentric (whereas ${\tilde\pi}^{ij}$ generate translations which do not preserve the positivity of the metric), because the Casimir invariant $T^AT^A$  is related to the kinetic operator in ${\bar H}$  through
\begin{equation}
\frac{\hbar^{2}\delta^{2}(0)}{2} T^{A}T^{A}={{\bar \pi}}^{i\dagger }_{j} {{\bar \pi}}^{j}_{i}= {{\bar \pi}}^{i}_{j}{{\bar \pi}}^{j}_{i} ={{\bar \pi}^{ij}}  {{\bar G}_{ijkl}} {{\bar \pi}}^{kl} .
\end{equation}
\noindent
The upshot is its spectrum can be labeled by eigenvalues of the complete commuting set at each spatial point comprising the two
Casimirs $L^{2}=T^{A}T^{A}, C=d_{ABC}T^AT^BT^C \propto \det({\bar \pi}^{i}_{j})$,  Cartan subalgebra $T^{3},T^{8}$, and isospin $I=\sum_{B=1}^3 T^{B}T^{B}$.   An underlying group structure has the advantage the action of momentric on wavefunctions  by functional differentiation can be traded for its well defined action as generators of $SU(3)$ on states expanded in this basis.  In addition, the ground state of the free theory, $|0\rangle$,  is thus an $SU(3)$  singlet state which is annihilated by all the momentric operators (i.e. $T^A|0\rangle =0$).

\section{Early universe and Cotton-York dominance}

As explained, early global intrinsic times and small volumes correspond to the era of Cotton-York dominance of ${\cal V}$ at the beginning of the universe,  wherein $\bar H = \sqrt{{\bar{\pi}}^{\dagger j}_i {\bar{\pi}}^{i}_j+g^2\hbar^2\tilde{C}^j_i\tilde{C}^i_j}$.
A number of intriguing facts conspire to simplify and regulate the Hamiltonian:
the  traceless  Cotton-York tensor density is expressible as ${\tilde C}^i_j =\bar{E}^i_{j(mn)}\frac{\delta W}{\delta \bar q_{mn}} $,
wherein $W=\frac{1}{4}\int{\tilde\epsilon}^{ijk}({\bar\Gamma}^l_{im} \partial_j{\bar\Gamma}^m_{kl} +\frac{2}{3}{\bar\Gamma}^l_{im}{\bar\Gamma}^m_{jn}{\bar\Gamma}^n_{kl})\,d^3x$ is the 3dDI Chern-Simons functional of the affine connection of ${\bar q}_{ij}$. This leads to the similarity transformation of the momentric,
\begin{equation}
{Q}^{i}_{j}= e^{gW}{{\bar \pi}}^{i}_{j}e^{-gW}=\frac{\hbar}{i}{\bar E}^i_{j(mn)}[\frac{\delta}{\delta{\bar q}_{mn}}- g\frac{\delta W}{\delta {\bar q}_{mn}}]=\frac{\hbar}{i}{\bar E}^i_{j(mn)}\frac{\delta}{\delta{\bar q}_{mn}}+ i g\hbar{\tilde C}^i_j .
\end{equation}
\noindent
Moreover, $[{\bar{\pi}}^i_j,\tilde{C}^j_i]=0$. Consequently, the Hamiltonian density is simply
$\bar H = {\sqrt {{Q}^{\dagger i}_{j}{Q}^{j}_{i}}}$\footnote{While ${Q}^{\dagger i}_{j}$ and ${Q}^{i}_{j}$ are related to ${\bar{\pi}}^i_j$ by $e^{\mp gW}$  similarity transformations, they are non-Hermitian, and generate two unitarily inequivalent representations of the non-compact group $SL(3,R)$ at each spatial point; whereas the momentric  ${\bar{\pi}}^i_j = \frac{1}{2}({Q}^{\dagger i}_{j} +{Q}^{i}_{j})$  generates a unitary representation of ${\prod}_{x }SU(3)_x$.}.

Ricci curvature terms become increasingly important in the potential after the initial era of Cotton-York dominance.
They can be introduced in a manner which preserves the underlying structure which regulate the Hamiltonian by extending the Chern-Simons action with
3dDI invariants of the spatial metric. This not only guarantees 3dDI invariance; but also makes the Hamiltonian density the square-root of a positive semi-definite and self-adjoint object
${Q}^{\dagger i}_{j}{Q}^{j}_{i}$; and ensures the preservation of all these properties even under renormalization of the coupling constants. In increasing order of spatial derivatives, these invariants are
$ \Lambda\int {\sqrt q} d^3x , EH =b\int {\sqrt q}R d^3x$, and the Chern-Simons functional of the affine connection with dimensionless coupling constant. Even higher derivative curvature invariants will come along with super-renormalizable dimensional coupling constants, while the cosmological constant volume term commutes with ${{\bar \pi}}^{i}_{j}$ due to the traceless projector $\bar{E}^i_{j(mn)}$. To wit, only the Einstein-Hilbert (EH) action in three dimensions and the Chern-Simons functional
 are relevant. This corresponds to adopting $W_T=\frac{g}{4}\int{\tilde\epsilon}^{ijk}({\bar\Gamma}^l_{im} \partial_j{\bar\Gamma}^m_{kl} +\frac{2}{3}{\bar\Gamma}^l_{im}{\bar\Gamma}^m_{jn}{\bar\Gamma}^n_{kl})\,d^3x + b\int {\sqrt q}Rd^3x$, which leads to
\begin{eqnarray}
&{Q}^{i}_{j}:= e^{W_T}{{\bar \pi}}^{i}_{j}e^{-W_T}=\frac{\hbar}{i}{\bar E}^i_{j(mn)}[\frac{\delta}{\delta{\bar q}_{mn}} - \frac{\delta W_T}{\delta{\bar q}_{mn}}]\nonumber\\&=\frac{\hbar}{i}{\bar E}^i_{j(mn)}\frac{\delta}{\delta{\bar q}_{mn}} +ib\hbar\sqrt{q}{\bar R}^i_j   + ig\hbar{\tilde C}^i_j;
\end{eqnarray}
wherein, again due to the ${\bar E}^i_{j(mn)}$ projector, only the traceless part of the Ricci tensor remains. The Hamiltonian density is then
\begin{eqnarray}
\bar H &=& {\sqrt {{Q}^{\dagger i}_{j}{Q}^{j}_{i}}} \nonumber\\
&=&\sqrt{{\bar{\pi}}^{\dagger j}_i {\bar{\pi}}^{i}_j+\hbar^2(g\tilde{C}^i_j + b\sqrt{q}{\bar R}^i_j)(g\tilde{C}^j_i +b\sqrt{q}{\bar R}^j_i ) +[{\bar{\pi}}^{i}_j, ib\hbar\sqrt{q}{\bar R}^j_i]}\,.
\end{eqnarray}
 Remarkably, the coincident commutator $[{\bar{\pi}}^{i}_j, ib\sqrt{q}\hbar{\bar R}^j_i] =-\frac{5}{12}b\hbar^2\delta(0){\sqrt q}(5{R}-\frac{9}{\epsilon})$\cite{ITQG}. Thus the potential for Einstein's theory, which is the spatial Ricci scalar and a (positive) c-number cosmological constant term, emerges. This means that, instead of the naive positive semi-definite form  with ${\cal V}$ as in (9),
 the  simple and elegant quantum Hamiltonian density  ${\sqrt {{Q}^{\dagger i}_{j}{Q}^{j}_{i}}}$ (with all its aforementioned advantages) already contains Einstein's GR with cosmological constant. Adopting this, the departures from Einstein's theory, which come from ${\bar R}^i_j$ and the Cotton-York tensor, only appear in the higher-curvature higher-derivative combination $(g\tilde{C}^j_i +b\sqrt{q}{\bar R}^j_i )(g\tilde{C}^i_j+b\sqrt{q}{\bar R}^i_j) $.  These `non-GR' terms are automatically absent in homogeneous FLRW cosmology, and also in constant curvature slicings of Painlev\'{e}-Gullstrand solutions of black holes\cite{Lin}. Consequently,  except for Cotton-York preponderance at very early times\footnote{Further discussions on the initial vacuum state, primordial fluctuations, relation to Penrose Weyl conjecture and the `arrow of time' can be found elsewhere \cite{ITQG}.}, Einstein's GR dominates at low curvatures and long wavelengths
in a theory in which `four-dimensional symmetry is not a fundamental property of the physical world'\cite{Dirac}.

In summary, the final Hamiltonian assumes the elegant form,
 \begin{equation}
 H_{Phys}={\hbar}\int \sqrt{(Q^A)^\dagger Q^A}\,\frac{\delta(0)}{{\sqrt 2}\beta}d^3x, \quad Q^A := e^{W_T}{T^A(x)}e^{-W_T};
  \end{equation}
  wherein  $\frac{\delta(0)}{\beta}d^3x$ is a dimensionless volume element, its divergence to be absorbed by renormalization of $\beta$ (renormalization of the  couplings constants in the theory remains to be  studied).  With dimensionless fundamental variables, the CR are\cite{ITQG,NCR}
  \begin{equation}
\label{RELATIONS}
\begin{split}
&[ \bar q_{ij}(x), \bar q_{kl}(y)]=0,\\
&[\bar q_{ij}(x), T^A(y)]= \frac{i}{2}\Big((\lambda^A)^k_i{\bar q}_{kj} + (\lambda^A)^k_j{\bar q}_{ki}\Big)\frac{\delta (x-y)}{\delta(0)};\\
&[T^{A}(x),T^{B}(y)]= i{f}^{AB}\,_CT^{C}\frac{\delta (x-y)}{\delta(0)}.
\end{split}
\end{equation}
 In conjunction with Schr\"{o}dinger evolution and time-ordering in global intrinsic time, this framework presents a new vista to resolve and surmount the many conceptual and technical challenges of quantum gravity.

\section*{Acknowledgments}

This work was supported in part by the Ministry of Science and Technology (R.O.C.) under Grant No. MOST104-2112-M-006-003.
I would like to thank Eyo Eyo Ita, Niall \'{O} Murchadha and Hoi-Lai Yu for beneficial discussions.

\end{document}